\documentstyle[amssymb,aps]{revtex}

\begin{document}
\title{The cosmological origin of time-asymmetry}
\author{Mario Castagnino}
\address{Instituto de Astronom\'\i a y F\'\i sica del Espacio\\
Casilla de Correos 67, Sucursal 28, 1428 Buenos Aires, Argentina}
\author{Luis Lara}
\address{Departamento de F\'\i sica, Universidad Nacional de Rosario\\
Av. Pellegrini 250, 2000 Rosario, Argentina}
\author{Olimpia Lombardi}
\address{CONICET - Universidad de Buenos Aires \\
Pu\'an 470, 1406 Buenos Aires, Argentina}
\maketitle

\begin{abstract}
In this paper we address the problem of the arrow of time from a
cosmological point of view, rejecting the traditional entropic approach that
defines the future direction of time as the direction of the entropy
increase: from our perspective, the arrow of time has a global origin and it
is an intrinsic, geometrical feature of space-time. Time orientability and
existence of a cosmic time are necessary conditions for defining an arrow of
time, which is manifested globally as the time-asymmetry of the universe as
a whole, and locally as a time-asymmetric energy flux. We also consider
arrows of time of different origins (quantum, electromagnetic,
thermodynamic, etc.) showing that they can be non-conventionally defined
only if the geometrical arrow is previously defined.
\end{abstract}

\section{Introduction}

In our previous papers on time-asymmetry (\cite{Cosmo}, \cite{CG},\cite
{Castagnino}, \cite{Goslar}, \cite{Peyresq}, \cite{Ordo}, \cite{cqg1}, \cite
{cqg2}), the main features of this problem were considered (e.g. the
thermodynamic arrow of time in early universe \cite{cqg1} or the arrow of
time in some cosmological models \cite{cqg2}). In this paper we will try to
present a comprehensive and updated formulation of the subject from a
cosmological point of view.

As we have said many times before, the problem of time-asymmetry, also known
as the problem of the arrow of time, can be formulated by means of the
following question: {\it How an evident time-asymmetry is possible if the
laws of physics are time-reversal invariant? }In fact, the laws of physics
are invariant under the transformation $t\rightarrow -t$\footnote{%
There are two exceptions:
\par
a.- The second law of thermodynamics: entropy grows. But we use to consider
this ''law'' as an empirical fact that must be demonstrated from more
primitive and elementary laws.
\par
b.- Weak interactions. But they are so weak that it is difficult to see how
the time-asymmetry of the universe can be derived from these interactions.
Therefore, as it is usual in the literature, we do not address this problem
in this paper.}. Nevertheless, we have the psychological feeling that past
is different than future; moreover, there are clear time-asymmetric
phenomena, being the natural tendency from non-equilibrium to equilibrium
the most conspicuous example. Astonishing enough, the solution is contained
in the above italized question. Since time-asymmetry cannot be explained by
the time-reversal invariant laws ({\it equations}) of physics, it should be
explained by some time-asymmetric {\it initial conditions.} But, at first
sight, initial conditions are arbitrary; therefore, it is impossible to
formulate a physical law on initial conditions. However, the initial
conditions of any process are produced by previous processes in such a way
that all processes in a connected universe are coordinated in some way.
Therefore, the reason of time-asymmetry is the asymmetry of the universe,
namely, a{\it \ global }reason. The aim of this paper is to explain this
fundamental fact in a way as conceptually clear as possible.

In the early nineteenth century origin of statistical mechanics, Boltzmann
asserted: ''The universe, or at least a big part of it around us, considered
as a mechanical system, began in a very improbable state and it is now also
in a very improbable state. Then if we take a smaller system of bodies, and
we isolate it instantaneously from the rest of the world, in principle this
system will be in an improbable state and, during the period of isolation,
it will evolve towards more probable states'' \cite{Boltz}. Since
Boltzmann's seminal work, many authors had the intuition that time-asymmetry
has a global origin. For instance, Feynman claimed: ''For some reason, the
universe at one time had a very low entropy for its energy content, and
since then entropy has increased. So that is the way towards future. That is
the origin of all\ irreversibility'' \cite{Feynman}. However, these
traditional positions define time-asymmetry in terms of entropy increase
(see also Davies \cite{Davies}). In the present paper we will reject the
traditional entropic approach, following John Earman's \cite{Ear} ''Time
Direction Heresy'' according to which the arrow of time is an intrinsic,
geometrical feature of space-time: this geometrical approach to the problem
of the arrow of time has conceptual priority over the entropic approach
since the geometrical properties of the universe are more basic than its
thermodynamic properties.

In order to develop our arguments, we will adopt two methodological
hypotheses which are unquestioningly accepted in present-day cosmology:

1.- The laws of physics are always and everywhere the same.

2.- The universe is unique: disconnected universes are not considered.

The paper is organized as follows:

In Section II the concepts of time-reversal invariance, irreversibility and
time-asymmetry will be distinguished. Section III is devoted to introduce
the difference between conventional and substantial arrows of time: this
will allow us to formulate the problem in precise terms. In Section IV we
will argue that the time-orientability of space-time and the existence of a
global time are necessary conditions for defining the arrow of time. On this
basis, we will develop a two steps program: Section V is devoted to show
that the universe is a time-asymmetric object; in Section VI we will argue
that this global time-asymmetry manifests itself in each point of the
manifold as a local time-asymmetry. In Sections VII and VIII we will
consider time-asymmetries of different origins (quantum, electromagnetic,
thermodynamic, etc.), namely, other ''arrows of time'': since the subject of
these sections is quite long but we would like to give a comprehensive
overview, in many cases we will refer the reader to the literature (mostly
to our own papers). In Section IX we will draw our conclusions. An Appendix
studying a 1999 paper of L. Schulman \cite{Schulman} completes this work:
the aim of this appendix is to demonstrate that the discussion on this
fundamental subject is still open.

\section{Time-reversal invariance, irreversibility, time-symmetry}

In general, these three concepts are invoked in the treatment of the problem
of the arrow of time, but usually with no elucidation of their precise
meanings; this results in confusions that contaminate many interesting
discussions. For this reason, we will start from providing some necessary
definitions:

1.- {\bf Time-reversal invariance: }It is a property of evolution equations
(laws) and, a fortiori, of the set of its solutions (models). An evolution
equation is {\it time-reversal invariant} if it is invariant under the
transformation $t\rightarrow -t$; as a result, for each solution $f(t)$, $%
f(-t)$ is also a solution.

2.- {\bf Reversibility: }It is a property of a single solution of an
evolution equation. A solution is {\it reversible} if it corresponds to a
closed curve in phase space.

Both properties can combine in the four possible alternatives:

a.- {\bf Time-reversal invariance and reversibility. }Let us consider the
harmonic oscillator with Hamiltonian: 
\[
H=\frac 12\,p^2+\frac 12\,K^2\,q^2 
\]
The equation is time-reversal invariant, namely, it is invariant under the
transformation $q\rightarrow q,$ $p\rightarrow -p$. As a result, the set of
trajectories is symmetric with respect to the $q$-axis. Since each
trajectory is a closed ellipse, it is reversible.

b.- {\bf Time-reversal invariance and irreversibility. }Let us consider the
pendulum with Hamiltonian: 
\[
H=\frac 12\,p_\theta ^2+\frac{K^2}2\,\cos \theta 
\]
Again the equation is time-reversal invariant, and the set of solutions is
symmetric with respect the $\theta $-axis. The trajectories within the
separatrices are reversible since they are closed curves. But the
trajectories above (below) the separatrices are irreversible since, in the
infinite time-limit, $\theta \rightarrow \infty $ ($\theta \rightarrow
-\infty )$. The trajectories in the upper (lower) separatrix are also
irreversible since they tend to $\theta =\frac \pi 2,p_\theta =0$ ($\theta =-%
\frac \pi 2,p_\theta =0$) when $t\rightarrow \infty $ $(t\longrightarrow
-\infty )$.

c.- {\bf Time-reversal non-invariance and reversibility. }Let us consider
the modified oscillator with Hamiltonian: 
\[
H=\frac 12\,p^2+\frac 12\,K(p)^2\,q^2 
\]
where $K(p)=K^{+}=const.$ {\tiny \ }when {\tiny \ }$p\geq 0$, and $%
K(p)=K^{-}=const.${\tiny \ }when {\tiny \ }$p<0.$ If $K^{+}\neq K^{-}$, the
equation is not time-reversal invariant: the set of solutions is clearly
asymmetric with respect to the $q$-axis. Nevertheless, each trajectory is
closed and, therefore, reversible.

d.- {\bf Time-reversal non-invariance and irreversibility. }Let us consider
a damped oscillator whose equation is: 
\[
\stackrel{\bullet \bullet }{q}=-K^2\,q-A^2\stackrel{\bullet }{q} 
\]
The equation is time-reversal non-invariant. The origin is an attractor and
the trajectories are spirals: they are trapped by the origin when $%
t\rightarrow \infty $. Therefore, each trajectory is irreversible.

Now we will introduce what we consider the most relevant concept in the
problem of the arrow of time:

3.- {\bf Time-symmetry}: It is a property of a single solution of an
evolution equation. A solution $f(t)$ is {\it time-symmetric} if there is a
time $t_S$ such that $f(t_S+t)=$ $f(t_S-t)$.

Since we will define the arrow of time as a global feature of the universe,
it is interesting to see how this definition of time-symmetry applies to
cosmology. Let us consider a FRW Big Bang-Big Crunch universe ($k=1$), where 
$a_{\max \text{ }}$ corresponds to a time $t_{\max }$; this universe is
time-symmetric if\footnote{%
Now we are just considering the geometry. Matter within the universe will be
considered in the next sections.}: 
\[
a(t_{\max }+t)=a(t_{\max }-t) 
\]
More generally, let us consider a universe endowed with a cosmic time $t$,
namely, with metric: 
\[
ds^2=dt^2-h_{ij}(t,x^k)\ dx^idx^j 
\]
We will say that this universe is time-symmetric if there is a time $t_{S%
\text{ }}$such that: 
\[
h_{ij}(t_S+t,x^i)=h_{ij}(t_S-t,x^i) 
\]

\section{Conventional vs. substantial arrows of time}

Traditional discussions around the problem of the arrow of time are usually
subsumed under the label ''the problem of the direction of time'', as if we
could find an exclusively physical criterion for singling out {\it the}
direction of time, identified with what we call ''the future''. But there is
nothing in (local) physics that distinguishes, in a non-arbitrary way,
between past and future as we conceive them. It might be objected that
physics implicitly assumes this distinction with the use of asymmetric
temporal expressions, like ''future light cone'', ''initial conditions'',
''increasing time'', and so on. However this is not the case, and the reason
can be understood on the basis of the distinction between the concepts of
conventional and substantial.

Two objects are {\it formally identical} when there is a permutation that
interchanges the objects but does not change the properties of the system to
which they belong or in whose description they are involved. In physics it
is usual to work with formally identical objects: the two lobes of a light
cone, the two spin senses, etc.

i.- We will say that we establish a {\it conventional} difference when we
call two formally identical objects with two different names, e.g., when we
assign different signs to the two spin senses.

ii.- We will say that the difference between two objects is {\it substantial}
when we give different names to two objects which are not formally identical
(see \cite{Penrose}, \cite{Sachs}). In this case, even though the names are
conventional, the difference is substantial. E.g. the difference between the
two poles of the theoretical model of a magnet is conventional since both
poles are formally identical; the difference between the two poles of the
Earth is substantial because in the north pole there is an ocean and in the
south pole there is a continent (and the difference between ocean and
continent remains substantial even if we conventionally change the names of
the poles).

Once this point is accepted, the problem cannot yet be posed in terms of
singling out the future direction of time: the problem of the arrow of time
becomes the problem of finding a {\it substantial difference} between the
two temporal directions. But if this is our central question, we cannot
project our independent intuitions about past and future for solving it
without begging the question. If we want to address the problem of the arrow
of time from a perspective purged of our temporal intuitions, we must avoid
the conclusions derived from subtly presupposing time-asymmetric notions. As
Huw Price \cite{HP} claims, it is necessary to stand at a point outside of
time, and thence to regard reality in atemporal terms. This atemporal
standpoint prevents us from using the asymmetric temporal expressions of our
natural language in a non-conventional way. But then, what does ''the arrow
of time'' mean when we accept this constraint? Of course, the traditional
expression coined by Eddington has only a metaphorical sense: its meaning
must be understood by analogy. We recognize the difference between the head
and the tail of an arrow on the basis of its geometrical properties;
therefore, we can substantially distinguish between both directions,
head-to-tail and tail-to-head, independently of our particular perspective.
Analogously, we will conceive the problem of the arrow of time in terms of 
{\it the possibility of establishing a substantial distinction between the
two directions of time on the basis of exclusively physical arguments}.

When the problem is formulated in these terms, it is clear that the solution
will consist in demonstrating the time-asymmetry of the universe: in a
time-asymmetric universe, the two directions of time can be substantially
distinguished. However, first it is necessary to explain the topological
conditions required to establish such a distinction.

\section{Time-orientability and cosmic time}

Earman \cite{Ear} and Gr\"unbaum \cite{Grun} were the first authors who
emphasized the relevance of time-orientability to the problem of the arrow
of time. In fact, general relativity considers the universe as a
pseudo-Riemannian manifold that may be time-orientable or not. {\it \ }A
space-time is {\it time-orientable} if and only if there exists a {\it %
continuous }non-vanishing time-like vector field globally defined. By means
of this field, the set of all light semi-cones (lobes) of the manifold can
be split into two equivalence classes, $C_{+}$ and $C_{-}$: the lobes of $%
C_{+}$ contain the vectors of the field and the lobes of $C_{-}$ do not
contain them. If space-time were not time-orientable, the distinction
between future lobes and past lobes would not be univocally definable on a
global level. On the other hand, in a time-orientable space-time, if there
were a time-reversal non-invariant law {\it L} defined in a {\it continuous}
way all over the manifold, this would allow us to choose one of the classes
as the future class (say $C_{+}$) and the other one as the past class (say $%
C_{-}$); the law {\it L} would be sufficient for defining the arrow of time
for the whole universe (namely, a future lobe $C_{+}(x)$ and a past lobe $%
C_{-}(x)$ at each point $x$). In fact, if one lobe of the class $C_{+}$ were
considered as the future lobe at a point $x$ and another lobe of the same
class were considered as the past lobe at a point $y$, then joining these
two points by means of a continuous curve (because we only consider
connected universes) and propagating the lobe of $x$ towards $y$ (and vice
versa) would be sufficient for finding a point where the law {\it L} would
be discontinuous, contrary to our assumption.

But, which is this global continuous time-reversal invariant law that allows
us to define past and future? This is the essence of Matthews' criticism 
\cite{Matt} to the relevance of time-orientability: since there are not
continuous and global time-reversal non-invariant laws of nature (but anyway
the arrow of time does exist), time-asymmetry is necessarily defined by
local laws and, then, it is just a local property; therefore, nothing rules
out the possibility that the arrow of time points to different directions in
different regions of space-time (see also Reichenbach \cite{Reichenbach}).
What Matthews has forgotten is that an asymmetric physical fact can be used
to define time-asymmetry instead of a physical law. Of course, it must be an
ubiquitous{\it \ }physical fact, because it must be used to define the
future and the past lobes at all the points of the universe. This ubiquitous
physical fact is the global time-asymmetry of the universe as a whole, which
manifests itself in all local domains (as we will explain in the next
sections).

However, in order to obtain the arrow of time, the possibility of defining a
cosmic time is a further requirement. A space-time has a {\it cosmic time}
if the following conditions hold:

i.- The space-time satisfies the {\it stable causality condition} \cite{HE}:
in this case, the manifold $M$ possesses a {\it global time function} $%
t:M\rightarrow {\Bbb R}$ whose gradient is everywhere time-like.

ii.- The global time function $t$ can be computed as the {\it distance}
between two hypersurfaces of the resulting foliation, measured along any
trajectory orthogonal to the foliation. In this case, the time function $t$
is the cosmic time.

Under the assumptions of time-orientability and existence of a cosmic time
(conditions satisfied by usual present-day cosmological models), in the
following sections we will show that:

1.- Time-symmetric universes belong to a set of {\it measure zero} on the
space of all possible universes\footnote{%
Precisely, we refer to the Liouville measure and to any measure absolutely
continuous with respect to it. In some cases it would more rigorous to say
that the dimension of the subspace of time-symmetric solutions of the
universe evolution is smaller than the dimension of the space of solutions,
meaning that time-asymmetric solutions are ''generic''.}.

2.- The global time-asymmetry of the universe can be used {\it locally }at
each point $x$ to define the future and the past lobes, $C_{+}(x)$ and $%
C_{-}(x)$.

We will develop these points in the next two sections.

\section{The corkscrew factory theorem}

\subsection{The theorem}

In his interesting book, Price \cite{HP} emphasizes that time-reversal
invariance is not an obstacle to construct a time-asymmetric model of the
universe: a time-reversal invariant equation may have time-asymmetric
solutions\footnote{%
Of course, this ''loophole'' is not helpful when we are dealing with a
multiplicity of systems: for each time-asymmetric solution there is another
time-asymmetric solution that is the temporal mirror image of the first one.
But when we are studying the whole universe, both solutions are equivalent
descriptions of one and the same universe.}. He illustrates this point with
the familiar analogy of a factory which produces equal numbers of
left-handed and right-handed corkscrews: the production as a whole is
completely unbiased, but each individual corkscrew is asymmetric. Price
argument shows the possibility of describing time-asymmetric universes by
means of time-reversal invariant laws. But, what is the reason to suppose
that time-asymmetric universes have high probability? We will demonstrate
that time-asymmetric solutions of the universe equations have measure zero
in the corresponding phase space.

Let us consider some model of the universe equations. All known examples
have the following two properties (e.g. see \cite{Lara}, but there are many
other examples):

1.- They are time-reversal invariant, namely, invariant under the
transformation $t\rightarrow -t.$

2.- They are time-translation invariant, namely, invariant under the
transformation $t\rightarrow t+const.$\footnote{%
We are referring to the equations that rule the behavior of the universe,
not to the particular solutions that normally do not have time-translation
symmetry.} (homogeneous time).

Let us consider the following cases:

a) The particular case of a FRW model where $a$ is the only dynamical
variable. It must satisfy the Hamiltonian constraint: 
\[
H(a,\stackrel{\bullet }{a})=0 
\]
namely, the Einstein equation. Then: 
\[
\stackrel{\bullet }{a}=f(a) 
\]
If $f(a)$ has a root at $a_S$ corresponding to the time $t_S$ and $\stackrel{%
\bullet }{f}(a_S)\neq 0$ (which would be a generic case), then at $t=t_S$
the radius $a$ has a maximum or a minimum. Since, by hypothesis, the
equations are time-reversal invariant and homogeneous (which allows us to
begin our reasoning from any point), both sides of $a(t)$ will be symmetric
with respect to $t_S$:

\[
a(t_S+t)=a(t_S-t) 
\]
Therefore, for this kind of universe global time-symmetry is generic%
\footnote{%
Essentially this was what Hawking called his ''greatest mistake''\cite{SH}.
In fact, time-symmetry appears to be generic only in the simplest
cosmological universes.} to the extent that the dimension of the set of
time-symmetric solutions is the same than the dimension of phase space.

b) Let us now consider a more generic case: a FRW universe with radius $a$
and matter represented by a neutral scalar field $\phi .$ The dynamical
variables are now $a,\stackrel{\bullet }{a},\phi ,\stackrel{\bullet }{\phi }$%
. They satisfy a generic Hamiltonian constraint:

\begin{equation}
H(a,\stackrel{\bullet }{a},\phi ,\stackrel{\bullet }{\phi })=0  \label{4.0}
\end{equation}
which reduces the dimension of phase space from 4 to 3; then, we can
consider a phase space of variables $\stackrel{\bullet }{a},\phi ,\stackrel{%
\bullet }{\phi }$ and: 
\begin{equation}
a=f(\stackrel{\bullet }{a},\phi ,\stackrel{\bullet }{\phi })  \label{4.1}
\end{equation}
a function obtained solving eq.(\ref{4.0}).

If we want to obtain a time-symmetric continuous\footnote{%
We will disregard non-continuous solutions since normally information do not
pass through discontinuities and we are only considering {\it connected}
universes where information can go from a point to any other timelike
connected point.} solution such that $a\geq 0$\footnote{%
As only $a^2$ appears in a FRW metric, we will consider just the case $a\geq
0$ since the point $a=0$ is actually a singularity that cuts the time
evolution.}, there must be a time $t_S$ regarding to which $a$ is symmetric:

\[
a(t_S+t)=a(t_S-t)\quad \text{ and\quad }\stackrel{\bullet }{a}(t_S)=0 
\]
In order to obtain complete time-symmetry, $\phi $ must also be symmetric
about $t_S$. There are two cases: even symmetry: 
\[
\phi (t_S+t)=\phi (t_S-t)\text{\quad and\quad }\stackrel{\bullet }{\phi }%
(t_S)=0 
\]
and odd symmetry: 
\[
\phi (t_S+t)=-\phi (t_S-t)\text{\quad and}\quad \phi (t_S)=0 
\]
This means that time-symmetric trajectories necessarily pass trough the axes 
$(0,\phi ,0)$ or ($0,0,\stackrel{\bullet }{\phi })$ of the phase space. From
these ''initial'' conditions we can propagate, using the evolution
equations, the corresponding trajectories; this operation will produce two
surfaces that contain the trajectories with at least one point of symmetry,
that is, that contain all the possible time-symmetric trajectories. Both
surfaces have dimension 2%
\mbox{$<$}
3 (namely, the dimension of our phase space). The usual Liouville measure of
these sets is zero, and also any measure absolutely continuous with respect
to it. In this way we have proved that, for generic models of the universe,
the solutions are time-asymmetric with the exception of a subset of
solutions of measure zero. q.e.d.\footnote{%
Of course, the models must be generic. E.g. if we chose an $H$ such that $%
\frac{\partial H}{\partial \phi }=\frac{\partial H}{\partial \stackrel{%
\bullet }{\phi }}=0$, we are in the case of point a) and time-symmetric
solutions become generic. But this is not a generic case. A more
(apparently) non-generic example could be obtained making the canonical
transformation of the model above.}.

\subsection{Generalization of the theorem}

The theorem can be easily generalized to the case where $\phi $ has many
components, or to the case of many fields with many components. Some of
these fields may be fluctuations of the metric: in this case, we must
Fourier transform the equations, and this would allow us to reproduce the
theorem only with $t$ functions. Since properties 1 and 2 (time-reversal
invariance and time-translation invariance) are also true in the classical
statistical case, the theorem can be also demonstrated in this case\footnote{%
When the phase space has infinite dimensions, it is better to use the notion
of dimension instead of that of measure, as explained in footnote 3.}. And
also in the quantum case, albeit some quantum gravity problems like time
definition\cite{PT}.

Let us now consider the coarse-grained version of the theorem. Let $%
\varepsilon $ be the size of the grain and, in order to compare measures,
let us consider that the phase space $\stackrel{\bullet }{a},\phi ,\stackrel{%
\bullet }{\phi }$ is a cube of volume $L^{3}.$ In this case, boundary
conditions $(0,\phi ,0)$ and ($0,0,\stackrel{\bullet }{\phi })$ will be
fuzzy and the volume of the set of time-symmetric initial conditions will
have measure $2\varepsilon ^{2}L.$ This magnitude can be compared with the
size of the phase space, obtaining the ratio $2\varepsilon
^{2}L/L^{3}=2(\varepsilon /L)^{2}.$ Of course, in the usual case $%
\varepsilon \ll L$; then, the measure of the set of points corresponding to
initial conditions that lead to time-symmetric universes is extremely
smaller than the measure of the phase space. The same argument can be
applied to the set of time-symmetric solutions with measure $2\varepsilon
L^{2}$, where $2\varepsilon L^{2}/L^{3}=2\varepsilon /L<<1$ if $\varepsilon
\ll L.$ q.e.d.

This completes the first argument announced in Section IV; let us now
develop the second argument.

\section{From global time-asymmetry to local time-asymmetry}

\subsection{The generic case}

Combining the results of Section IV and Section V it is not difficult to
achieve the second step of our program.

>From Section IV we know that if a space-time is time-orientable, a
continuous non-vanishing time-like vector field $\gamma ^\mu (x)$ can be
defined all over the manifold. At this stage, the universe is {\it %
time-orientable} but not yet {\it time-oriented}, because the distinction
between $\gamma ^\mu (x)$ and $-\gamma ^\mu (x)$ is just conventional. Now
Section V comes into play. A time-orientable space-time having a cosmic time 
$t$ is time-asymmetric if there is not a time $t_S$ that splits the manifold
into two ''halves'', one the temporal mirror image of the other regarding
their intrinsic geometrical properties. This means that, in a
time-asymmetric universe, any time $t_A$ splits the manifold into two
sections that are different to each other: the section $t>t_A$ is {\it %
substantially} different than the section $t<t_A$. We can chose any point $%
x_0$ with $t=t_A$ and conventionally consider that $-\gamma ^\mu (x_0)$
points towards $t<t_A$ and $\gamma ^\mu (x_0)$ points towards $t>t_A$ or
vice versa: in any case we have established a substantial difference between 
$\gamma ^\mu (x_0)$ and $-\gamma ^\mu (x_0)$. We can conventionally call
''future'' the direction of $\gamma ^\mu (x_0)$ and ''past'' the direction
of $-\gamma ^\mu (x_0)$ or vice versa, but in any case past is substantially
different than future. Now we can extend this difference to the whole
continuous fields $\gamma ^\mu (x)$ and $-\gamma ^\mu (x)$: in this way, the
time-orientation of the space-time has been established. Since the field $%
\gamma ^\mu (x)$ is defined all over the manifold, it can be used {\it %
locally }at each point $x$ to define the future and the past lobes: for
instance, if we have called ''future'' the direction of $\gamma ^\mu (x)$, $%
C_{+}(x)$ contains $\gamma ^\mu (x)$ and $C_{-}(x)$ contains $-\gamma ^\mu
(x)$. This is the solution of the second step in the general case.

\subsection{From the generic case towards our own universe.}

Even if the second step has been completed in the generic case, it is clear
that the solution is more mathematical than physical. Thus, it would be
desirable to show how the general time-orientation is reflected in everyday
physics, where time-asymmetry manifests itself in terms of time-asymmetric
energy fluxes. However, this task will lead us to impose reasonable
restrictions in the considered cosmological model in such a way that the
explanation of local time-asymmetry applies, not to the generic case, but
rather to the particular case of our own universe.

i.- Up to this point, global time-asymmetry has been considered as a
substantial asymmetry of the geometry of the universe, embodied in the
metric tensor defined at each point of the space-time: $g_{\mu \nu }(x)$.
Perhaps the easiest way to see how this geometrical time-asymmetry is
translated into local physical terms is to consider the energy-momentum
tensor, which can be computed by using $g_{\mu \nu }(x)$ and its derivatives
through Einstein's equation: 
\[
T_{\mu \nu }=\frac 1{8\pi }\,\left( R_{\mu \nu }(g)-\frac 12\,g_{\mu \nu
}\,R(g)-\Lambda \,g_{\mu \nu }\right) 
\]
The curvatures $R_{\mu \nu }(g)$, $R(g)$ can be obtained from $g_{\mu \nu
}(x)$ and its derivatives, and $\Lambda $ is the cosmological constant. Now
we impose a first condition: that our $T_{\mu \nu }$ turns out to be a ''%
{\it normal}'' or {\it Type I} energy-momentum tensor. Then, $T_{\mu \nu }$
can be written as: 
\[
T_{\mu \nu }=s_0\,V_\mu ^{(0)}\,V_\nu ^{(0)}+\sum_{i=1}^3\,s_i\,V_\mu
^{(i)}\,V_\nu ^{(i)}
\]
where $\left\{ V_\mu ^{(0)},\,V_\mu ^{(i)}\right\} $ is an orthonormal
tetrad, $V_\mu ^{(0)}$ is time-like and the $V_\mu ^{(i)}$ are space-like ($%
i=1,2,3$) (see \cite{HE}, \cite{Lichne}). Since we have assumed that the
manifold is continuous, $g_{\mu \nu }(x)$ and $T_{\mu \nu }(x)$ are
continuously defined over the manifold (provided the derivatives of $g_{\mu
\nu }(x)$ are also continuous); this means that $V_\mu ^{(0)}(x)$ is a
continuous unitary time-like vector field defined all over the manifold,
which can play the role of the field $\gamma ^\mu (x)$ if everywhere $%
s_0\neq 0$ ($V_\mu ^{(0)}$, even if time-like, may change its sign when $%
s_0=0)$.

Here we impose a second condition: that the universe satisfies the {\it %
dominant energy condition: }i.e. $T^{00}\geq $ $\left| T^{\mu \nu }\right| $
in any orthonormal basis (namely, $s_0\geq 0$ and $s_i\in \left[
-s_0,s_0\right] $). In this case, $s_0\neq 0$ and, then, $V_\mu ^{(0)}(x)$
is continuous, time-like and non-vanishing. This means that $V_\mu ^{(0)}(x)$
can play the role of $\gamma ^\mu (x)$, with the advantage that it has a
relevant physical sense. In this way, a time-orientation is chosen at each
point $x$ of the manifold, and the time components of $T_{\mu \nu }$ acquire
definite signs according to this orientation. We will use this orientation
in our arguments below. Therefore we have translated the global
time-asymmetry into local terms, endowing the local arrow with a physical
sense.

ii.- Since we are now in local grounds, our new task is to understand the 
{\it local nature }of the characters in the play. If $T^{00}\geq $ $\left|
T^{\mu \nu }\right| $, then $T^{00}\geq $ $\left| T^{i0}\right| $.
Therefore, $T^{0\mu }$, which is usually considered as the local energy
flux, is a time-like (or light-like) vector. This holds for all presently
known forms of energy-matter and, so, there are in fact good reasons for
believing that this should be the case in almost all situations (for the
exceptions, see \cite{MV}\footnote{%
E.g., some exceptions are: Casimir effect, squeed vacuum, Hawking
evaporation, Hartle-Hawking vacuum, negative cosmological constant, etc.
These objects are strange enough in nowadays observational universe to
exclude the practical existence of zones with $T^{00}$ of different signs
and, therefore, with different time directions (see also \cite{BV}).}).

iii.-But, is really $T^{0\mu }$ the energy flux? To go even closer to
everyday physics, we must remember that $T_{\mu \nu }$ satisfy the
''conservation'' equation: 
\[
\nabla _\mu \,T^{\mu \nu }=0 
\]
Nevertheless, as it is well known, this is not a true conservation equation
since $\nabla _\mu $ is a covariant derivative. The usual conservation
equation with ordinary derivative reads: 
\[
\partial _\mu \,\tau ^{\mu \nu }=0 
\]
where $\tau _{\mu \nu }$ is not a tensor and it is defined as: 
\[
\tau _{\mu \nu }=\sqrt{-g}\ (T_{\mu \nu }+t_{\mu \nu }) 
\]
where we have introduced a $t_{\mu \nu }$ that reads: 
\[
\sqrt{-g}\ t_{\mu \nu }=\frac 1{16\pi }\left[ {\cal L}\,g_{\mu \nu }-\frac{%
\partial {\cal L}}{\partial g_{\mu \nu },\lambda }\ g_{\mu \nu },\lambda
\right] 
\]
where ${\cal L}$ is the Lagrangian. $t_{\mu \nu }$ is also an homogeneous
and quadratic function of the connection $\Gamma _{\nu \mu }^\lambda $ \cite
{LL}. Now we can consider the coordinates $\tau ^{0\mu }$, which satisfy: 
\[
\partial _\mu \tau ^{0\mu }=\partial _0\tau ^{00}+\partial _i\tau ^{0i}=0 
\]
namely, a usual conservation equation. Even if $\tau ^{0\mu }$ is not a
four-vector, it can be defined in each coordinate system: in each system $%
\tau ^{00}$ can be considered as the energy density and $\tau ^{0i}$ as the
energy flux (the Poynting vector). This means that the field $\tau ^{0\mu
}(x)$ represents the spatio-temporal energy flow within the universe better
than $T^{0\mu }$.

In particular, in a local inertial frame where $\Gamma _{\nu \mu }^\lambda
=0 $, we have $\tau _{\mu \nu }=\sqrt{-g}\ T_{\mu \nu }$: in orthonormal
coordinates, the dominant energy condition will be now $\tau ^{00}\geq $ $%
\left| \tau ^{i0}\right| $ and $\tau ^{0\mu }$ will be time-like (or
light-like). But $\tau ^{0\mu }$ is just a local energy flow since it is
defined in orthonormal local inertial frames. Nevertheless, in any moving
frame with respect to those ones, if the acceleration of the moving frame is
not very large, the $(\Gamma _{\nu \mu }^\lambda )^2$ and the $t_{\mu \nu }$
are very small and the energy flux in the moving frame is time-like (or
light-like) for all practical purposes. This is precisely the case of the
commoving frame of our present-day universe: the arrows of Fig.1 represent
the $\tau ^{0\mu }$ in this case.

In summary, $\tau ^{0\mu }$ (that can locally be considered as the
four-velocity of a quantum of energy carrying a message) is a time-like {\it %
local} energy flux and:

a) It inherits the global time-asymmetry of $g_{\mu \nu }(x)$, i.e., the
geometrical time-asymmetry of the universe.

b) It translates the global time-asymmetry into the local level: the lobes $%
C_{-}(x)$ receive an incoming flux of energy while the lobes $C_{+}(x)$ emit
an outgoing flux of energy.

iv.- In order to complete the argument, we should consider all possible
universes, that is, not only our (Big Bang-Big Chill) universe, but also all
conjectural universes (satisfying the conditions required above). Of course,
in practice this is an impossible task to the extent that we do not know the
phenomenology of these conjectural universes. Therefore, in the next
subsection we will continue the analysis of the energy flux just in the
universe we inhabit.

\subsection{Particular case: our own universe. The Reichenbach-Davies diagram
}

In his classical book about the arrow of time, Hans Reichenbach \cite
{Reichenbach} defines the future direction of time as the direction of the
entropy increase of the majority of branch systems, that is, systems which
become isolated or quasi-isolated from the main system during certain
period. Paul Davies \cite{Davies} appeals to Reichenbach's notion, claiming
that branch systems emerge as the result of a chain or hierarchy of
branchings which expand out into wider and wider regions of the universe;
therefore ''the origin of the arrow of time refers back to the cosmological
initial conditions''.

On the basis of this idea, in previous papers we have introduced the
''Reichenbach-Davies diagram'' \cite{Peyresq}, \cite{cqg1}, \cite{CastKL}%
\footnote{%
At the classical level, the ''Reichenbach-Davies diagram'' can be considered
as the combination of all the classical scattering processes within the
universe. We have called ''Reichenbach-Bohm diagram'' to the combination of
all the quantum scattering processes within the universe, i.e. the quantum
version of the ''Reichenbach-Davies diagram'' \cite{Goslar}, \cite{Ordo}.},
where all the local processes which go from non-equilibrium to equilibrium
are connected in such a way that the ''output'' of a process is the
''input'' of another one: the energy provided by a process relaxing to
equilibrium serves to drive another process to non-equilibrium. This
''cascade'' of processes defines a global energy flux in our universe which,
if traced back, owes its origin to the initial global instability that is
the source of all the energy of the universe. The existence of the initial
global instability can be deduced from the equations of Section V. If we
consider that the universe begins in a Big Bang with $a=0,$ from eq.(\ref
{4.1}) all possible Big Bang initial conditions are contained in the surface:

\[
f(\stackrel{\bullet }{a},\phi ,\stackrel{\bullet }{\phi })=0 
\]
of phase space. The generic points of this surface are unstable (i.e. they
are not attractors); then, a generic Big Bang beginning is unstable. Of
course, the unstable nature of the solutions of the generic Einstein
equation is a well known fact. Moreover, using entropic considerations the
initial instability is studied in \cite{Aqui} and bibliography therein.
Quantum universes are also unstable, e.g. Vilenkin universe, which is
essentially a quantum system with a potential barrier allowing quantum
tunneling\cite{AV}. Another unstable quantum model is studied in \cite{MAC},
etc. This means that the existence of the initial instability is well
established.

The Reichenbach-Davies global system is the system of all time-asymmetric
processes within the universe: any process of the system begins in an
unstable state that was produced using energy coming from another process of
the global system\footnote{%
Since Reichenbach\cite{Reichenbach} does not take into account the
time-orientability of space-time, he accepts the possibility of a universe
with no global arrow of time, with regions having arrows ''pointing'' to
opposite directions. However, this possibility not only can be objected on
theoretical basis (see Section IV), but also seems unplausible on
observational grounds: nowadays we know that visible universe has a unique
arrow of time, since there is no astronomical observation showing that the
time-asymmetric behavior of nature would be inverted in some (eventually
very distant) regions of the universe. In fact, supernovae evolutions always
follow the same pattern (form birth to death, like human beings !!!), and
there is no trace of an inverted pattern in all the universe. This is a
relevant observation when we consider that supernovae are the markers used
to measure the longest distances corresponding to objects near the limit of
the visible universe \cite{Dubner}.}. In fact, energy always comes from
unstable$\rightarrow $stable or non-equilibrium$\rightarrow $equilibrium
processes (coal burning into ashes, H burning into He, etc.). The global
system (symbolized in Fig.1) has a{\it \ }time-asymmetry that we may call ''%
{\it global arrow of time}'' (GAT): this arrow points in opposite direction
to the initial cosmological instability and follows the evolution of all the
hierarchical chain towards equilibrium\footnote{%
At least in a expanding universe (Big Bang-Big Chill) case, which seems to
be the case of our universe.}. Each system in the diagram is called a {\it %
branch system} and it is represented by a box. The arrows coming from the
left of each box represent the energy produced by other boxes: part of the
energy going to the right is used to produce new unstable states, and the
rest is degraded (the degraded energy is represented by the outgoing arrows
that do not end in a box). This is also an asymmetry of the diagram of
Fig.1: the arrows corresponding to degraded energy only appear at the right
of each box. This is much more than just a detail, since it means that we
have a concentrated source of energy in the extremity of the universe that
we call ''the past'', from which we can pump energy to create other
concentrated sources of energy, while energy gets diluted towards what we
call ''the future''. In fact, this is the case of realistic Big Bang-Big
Chill cosmological models: energy is concentrated towards the past and
diluted towards the future \cite{MdU}, and this is another manifestation of
the time-asymmetry of the universe.

The global system also allows us to introduce the notion of {\it causality }%
in the universe (i.e. {\it global causality)} since, using Fig.1, we can say
that events A and B are not causally related, while C is the (partial) cause
of D, and D is the (partial) effect of C. On this basis we can say that {\it %
no effect can occur before its cause: }this statement is meaningful because
we have a global time-asymmetry that defines the word ''before''. The
physical substratum of causality is the energy coming from an unstable state
and creating a new unstable state; unstable structures are created by
pumping energy from sources in the past and decay spontaneously to
equilibrium towards the future.

Now we can go back to the manifold description of the previous subsection.
As we have seen, the time-asymmetry of the global energy flux is a
consequence of the time-asymmetry of the geometry of the universe. The
direction of the energy flux on a time-orientable space-time defines a
global {\it time-orientation}: the incoming flux defines the lobe $%
C_{-}(x)\in C_{-}$ at each point $x$, the outgoing flux defines the lobe $%
C_{+}(x)\in C_{+}$ at $x$, and all the lobes of class $C_{-}$ point towards
the initial instability. In this way, the global time-asymmetry of the
universe defines the local time-asymmetry in each one of its points. This
means that the energy flux is {\it the ubiquitous phenomenon} that locally
defines the arrow of time, because it can be found everywhere in the
universe. If two sections of the universe are not connected by such a flux,
then they are completely isolated from each other, and each one of them can
be conceived as a universe by itself; but this situation is not considered,
according to the second methodological hypothesis of the Introduction. The
global Reichenbach-Davies diagram defines the arrow of time of our universe
and, in this scenario, we can see how unstable states reach equilibrium
becoming stable states, and how entropy grows since it grows from unstable
to stable states. Therefore, the different ''arrows of time'' (cosmological,
thermodynamic, quantum, electromagnetic, etc.) can be coordinated, as we
will see in the next sections (see also \cite{Peyresq}).

\section{Other classical arrows of time}

The time-asymmetry studied in the previous sections reappears in the
phenomena explained by many chapters of physics, giving rise to many
''arrows of time'' corresponding to the different chapters, e.g.: the
cosmological arrow of time (CAT), the electromagnetic arrow of time (EMAT),
the quantum arrow of time (QAT), the thermodynamic arrow of time (TAT), etc.
They all result from the global time-asymmetry (GAT) and, therefore, point
to the same direction. In this section we will only give a schematic
introduction to this subject, since its full treatment exceeds the limits of
this paper; however, it is necessary to mention these points in order to
supply a complete account of the problem of the arrow of time. For each
arrow (with the exception of CAT) we will show that the time-reversal
invariant equations of evolution always produce two mathematical structures
symmetrically related by a time-reversal transformation (that we will call ''%
{\it t-symmetric twins}''), which usually embody notions related with
irreversibility. However, at this level the two structures are only
conventionally different: the problem here consists in supplying a
non-arbitrary criterion for choosing one of them. Only GAT allows us to
select one of these structures (one of the twins) as the one related to the
future or as the physically relevant one for spontaneous evolutions, by
creating a substantial difference between them.

\subsection{Cosmological arrow of time}

CAT is embodied in the fact that the radius of the universe grows, using the
direction of time defined by GAT (if not, the sentence would be
meaningless). Then, CAT points to the same direction than GAT in expanding
universes like ours. The usual irreversible models of present-day cosmology (%
$a\sim t^{\frac 12}$ or $a\sim t^{\frac 23}$ or $a\sim e^{Ht}$) are all
growing, and recent cosmological observations show that this is the case for
our actual universe \cite{BG}. Let us observe that even if CAT changes its
direction when a (conjectural) universe passes from an expanding phase to a
contracting phase, GAT never changes since it is defined by the global
time-orientation of the universe-manifold, as explained in Section VI.

\subsection{Electromagnetic arrow of time}

i.- EMAT is embodied in the fact that we must use retarded solutions in
electromagnetic problems. But electromagnetism provides us a pair of
advanced and retarded solutions (the pair of t-symmetric twins corresponding
to this case) which are only conventionally different. These solutions are
related also with the incoming and outgoing states in scattering situations.
The mathematical structure of these incoming and outgoing states is
rigorously defined by the Lax-Phillips scattering theory \cite{LP}.

ii.- In the Reichenbach-Davies diagram of Fig.1 we see that all spontaneous
evolutions (namely, the arrows going towards the right side of the diagram)
are outgoing states from the corresponding boxes, and these outgoing
(spontaneously evolving) states correspond to retarded solutions\footnote{%
EMAT can also be explained on cosmological grounds by the existence of an
absorbing black body at decoupling time \cite{Zeh}. Moreover, in
Lax-Phillips theory outgoing states correspond to retarded solutions and to
Hardy classes from below, anticipating the results of Subsection VIII.B.}.
The incoming states (the arrows going into the boxes coming from the left
side) are not spontaneous evolutions since they pump energy from the past to
create unstable states, and correspond to advanced solutions. Therefore, the
Reichenbach-Davies diagram, which defines the past and the future directions
of time, generates also a substantial difference between both members of the
pair of t-symmetric twins: retarded solutions are obtained by means of
energy coming from the past, whereas advanced solutions are obtained by
means of energy coming from the future. Since the global energy flux comes
from the past initial instability, only retarded solutions are physically
meaningful: this fact defines EMAT which, so defined, coincides with GAT.

\subsection{Thermodynamic arrow of time}

i.-TAT corresponds to the entropy increase in spontaneous evolutions of
closed systems, as prescribed by the second law of thermodynamics. It is
well known that many processes within the universe are chaotic: the
existence of chaotic mixing systems (like the well known examples of the
Gibbs ink drop, the sugar lump or the bottle of perfume) is an obvious fact.
Chaotic subsystems within the universe make the universe as a whole a
chaotic system since some of its parts are chaotic. For spontaneous
evolutions of chaotic closed systems an entropy can be defined \cite{Mackey}%
, {\it which monotonically increases from non-equilibrium in the present to
equilibrium in the future}. But, if starting from non-equilibrium in the
present we compute the entropy evolution towards the past, we will see that
entropy also grows in this time direction. This fact was already pointed out
by Paul and Tatiana Ehrenfest \cite{Ehrenfest} in their criticisms to Gibbs'
approach (see also \cite{Ruso}). The two temporally opposed evolutions are
the pair of t-symmetric twins corresponding to this case. The difference
between the twins is just conventional, and both express the irreversible
nature of thermodynamic processes going towards equilibrium either in the
past or in the future.

ii.- To break this symmetry we must know which one of both processes
corresponds to the spontaneous evolution, as we explained at the end of the
previous section. Using this criterion we can know which is the spontaneous
evolution and which is the non-spontaneous one. Then we can meaningfully say
that entropy grows in the spontaneous evolutions of closed systems. So GAT
gives rise again to a substantial difference between the two members of the
pair. If we define the ''total entropy'' of the matter-energy within the
universe\footnote{%
We are not considering an eventual entropy of the gravitational field.} by
adding the entropies of the {\it spontaneous evolutions} of all the closed
subsystems of the universe, since all these entropies increase {\it from
non-equilibrium to equilibrium}, the total entropy of the universe increases
in the same direction, from the initial unstable state to the final
equilibrium state. Therefore, TAT points to the same direction as GAT%
\footnote{%
Here a simple example of a Lyapunov variable could be in order. Let us
consider the typical and polemical case where $a$ has a maximum at some $t$
and begins and ends as $a=0$. As in eq.\ (\ref{4.1}) we can have 
\[
\stackrel{\bullet }{a}=-L(a,\phi ,\stackrel{\bullet }{\phi }) 
\]
and $\stackrel{\bullet \bullet }{a}\leq 0$ for all the evolution. So $L$ is
a Lyapunov evergrowing variable that depends on matter (trough $\phi ,%
\stackrel{\bullet }{\phi })$ and on geometry (trough $a)$, as we would
expect. This trivial fact shows how easy it is to find Lyapunov variables as
soon as the universe has just small complexity. Another example is the
''entropy gap'', as studied in paper \cite{Aqui} and bibliography therein.}.
Of course, we may use as well other ways of defining the ''total entropy''
of the universe, e.g.: we can define directly a conditional entropy and then
find that this entropy increases at least during the period where the
universe is close to thermal equilibrium \cite{Aqui}, or we can consider the
entropy produced by the creation of particles in the early universe \cite
{cqg1} and obtain that it also increases.

\subsection{Chaotic processes (from classical probabilities to classical
facts)}

Before we roll a (classical) dice, we can only foresee that the probability
for each face to come up is 1/6 (this is a classical probabilistic prophesy,
quantum ones will be treated in Subsection VIII.C) but, of course, we cannot
say which is the face that will come up. After the dice is rolled, we
certainly know which face is up: this a classical fact and such facts
produce a classical history. We must explain why this evolution $%
probabilities\longrightarrow facts$ (or if you prefer $prophesy\rightarrow
history)$ occurs towards the future and not the toward the past.

i.-The dice is laying on the table at time $-T$ and we know exactly which
one of its faces is up. Then, it undergoes a chaotic motion (in a dice cup
and a gambling table). Finally it remains at rest on the table at time $T$,
and we know again which one of its faces is up. This means that, under this
description (even though not in all its details) the phenomenon is
time-symmetric: we are certain at $t=\pm T$ and we can state just
probabilities at, let us say, $t=0$. This phenomenon can be decomposed in a
pair of t-symmetric twins: the process from $t=-T$ to $t=0$ and the process
from $t=0$ to $t=T$. Again, the problem consists in supplying a
non-conventional difference between both processes, that is, in explaining
why we only consider the process $t=0\rightarrow t=T$ and never the process $%
t=-T\rightarrow t=0$ when we gamble\footnote{%
All aleatory phenomena in nature can be analyzed as we have done with the
dice. E.g. the water in the sea is in an equilibrium state. The energy
coming from the sun (that we can consider as a source of energy where H is
transformed in He and, therefore, can be placed in the Reichenbach-Davies
diagram) evaporates this water. Now the same water belongs to a
meteorological chaotic system whose state can only be approximately foreseen
(with only 4 or 5 days of anticipation). Then, rain falls and the same water
ends in the sea, in an equilibrium state as initially.}.

ii.- The explanation begins by considering the states at $t=\pm T$ as
equilibrium states and the state at $t=0$ as a state out of equilibrium. In
fact, the dice was on the table with some face up at $t=-T$ . We take the
dice out of this equilibrium position at $t=0$ by using energy (from a box
of the Reichenbach-Davies diagram), and put it in chaotic motion (if we do
not use energy, the dice will stay forever in its equilibrium position).
Then, at $t=T$ the dice ends in its final equilibrium state. The first
process is forced, since we have pumped energy from the past; the second
process is spontaneous, since the dice returns spontaneously to a
equilibrium position. So the global asymmetry of the Reichenbach-Diagram
(i.e. GAT) introduces the desired substantial difference between the forced
evolution $t=-T\rightarrow t=0$ that belongs to $C_{-}(x)$ and the
spontaneous evolution $t=0\rightarrow t=T$ that belongs to $C_{+}(x)$.

Now we understand why classical probabilities (prophesy) always comes before
than classical facts (history). We can state this subject in another way:
the Reichenbach-Davies diagram allows us to explain the history of an
observer along his/her time-like path as the increase of his/her amount of
information \cite{CG}. The main fact is that messages are transported by
means of energy originated in unstable states that evolve towards
equilibrium. Therefore, the flow of messages follows the universal flow of
energy described in Section VI, namely, from past to future. This means that
the information increase of an observer can be illustrated as in Fig.2,
where the central curve symbolizes the observer space-time path and the
curves arriving to this path are the messages carried by energy coming from
unstable states (symbolized by boxes). We can then see that the amount of
information of the observer increases from the left side to the right side
of the diagram, creating his/her history.

\section{The quantum arrow of time}

\subsection{Quantum measurement arrow of time.}

Let us consider the simple example of a measurement process in the
scattering experiment of Fig.1 (dotted box). If we want to describe the
complete preparation-measurement process, we must consider not only the
scattering process itself, but also the accelerator that prepares the beam,
and the measurement apparatus, namely, the detector. The accelerator obtains
its energy from a source, where a decaying process takes place. In the
detector, a creation of an unstable state and a decaying process occur,
e.g., the particles to be detected are counted by a Geiger counter, where
they interact with a gas whose states are first excited (creation process)
and then decay (the energy obtained from this decaying process is used to
count the passing particles). The complete process of
preparation-measurement corresponds to the dotted box of Fig.1, that we
reproduce in Fig.3.

i.- Nevertheless, if we do not consider the origin or the fate of the
energies related to the preparation-measurement process (i.e. the direction
of the flux of energy), we have no substantial criterion for distinguishing
between the original scattering process and its temporal mirror image: this
is another example of t-symmetric twins.

ii.-But every preparation-measurement process takes place within the
Reichenbach-Davies system, since the energy comes from a source of energy
that can only be found in this global system. Therefore, the process that
goes from preparation to measurement turns out to be essentially different
from the time-reversed one: since preparation needs the energy that comes
from the hierarchical chain, it is the first process; the measurement is a
decaying process that produces degraded energy flowing towards the future.
Since QAT goes from {\it preparation} to {\it measurement }\cite{Ludwig},
its direction agrees with the direction of GAT.

\subsection{Minimal irreversible quantum mechanics}

The time-asymmetry of the universe also leads to the possibility of
formulate an irreversible quantum mechanics \cite{BohmPR}, \cite{Sudarshan}
or, more precisely, a {\it minimal} irreversible quantum mechanics \cite{CG}%
, \cite{Castagnino}, \cite{Ordo}, \cite{LC}.

i.- In fact, when we make the analytical extension of the energy spectrum of
the quantum system's Hamiltonian into the complex plane, we find poles in
the lower half-plane corresponding to decaying unstable states, and also
symmetric poles in the upper half-plane corresponding to growing unstable
states. These poles obviously correspond to irreversible processes.
Moreover, the symmetric position of these poles shows the time-reversal
invariance of the evolution equation: these pairs of poles can be considered
as the best illustration of pairs of t-symmetric twins.

ii.- But the growing states are created by the energy pumped by previous
unstable states, and the decaying states provide energy for latter
processes: the first energy flux is contained in $C_{-}(x)$ and the second
in $C_{+}(x)$. Therefore, the time-asymmetry of the universe defines the
substantial difference between growing and decaying unstable states, and
allows us to distinguish which poles are growing and which poles are
decaying. A cosmological description of this fact is given in \cite{MAC}.

Moreover, in order to transform the poles in Gamov vectors it is necessary
to introduce two subspaces of the Hilbert space ${\cal H}$, which we will
call $\phi _{\pm }$ \cite{Castagnino}. The states $|\varphi \rangle $ of the
subspace $\phi _{+}$ ($\phi _{-}$) are characterized by the fact that their
projections $\langle \omega |\varphi \rangle $ (on the energy eigenstates $%
|\omega \rangle $) are functions of the Hardy class from above (below). We
can prove that incoming quantum states belong to the Hardy class from above
and outgoing quantum states belong to the Hardy class from below, being this
fact the basis to understand decaying and growing processes and, therefore,
also the basis of irreversible quantum mechanics (see \cite{Ordo}). But, up
to this point, the Hilbert space ${\cal H}$ is time-reversal invariant in
the sense that: 
\[
K{\cal H=H} 
\]
where $K$ is the Wigner antilinear time-inversion operator. However, spaces $%
\phi _{\pm }$ are not time-reversal invariant since: 
\[
K\phi _{\pm }=\phi _{\mp } 
\]
Then, Hilbert space is a space that cannot be used to formulate a
time-reversal non-invariant quantum mechanics \cite{Peyresq}; the
substitution ${\cal H\rightarrow }\phi _{-}$ or ${\cal H\rightarrow }\phi
_{+}$ is the minimal modification that we should introduce into quantum
mechanics in order to make the theory time-reversal non-invariant.
Nevertheless, up to this point the difference between $\phi _{+}$ and $\phi
_{-}$ is just conventional (they are t-symmetric twins). But we can
establish a substantial difference if we use GAT to relate the incoming
states with the flux of energy coming from the initial global instability
and the outgoing states with the flux of energy that goes to the final
equilibrium state of the universe. This amounts to consider the space $\phi
_{-}$ as the physically meaningful and, therefore, to retain the lower poles
corresponding to the spontaneous decaying processes belonging to $C_{+}(x)$.

\subsection{Decoherence and the classical limit}

After this two introductory subsections, we would like to study the
so-called classical limit (i.e. the essential ingredient of the measurement
process) in detail. This limit can be conceptually analyzed in two steps.
The first step turns quantum mechanics into classical statistical mechanics;
in this phase we deal with probabilities. In the second step we go from
classical statistical mechanics to classical mechanics. The classical limit
is one of the phenomena that more eloquently shows time-asymmetry since this
limit always occurs towards the future.

Let us study the two steps in detail.

\subsubsection{Quantum mechanics$\rightarrow $classical statistical mechanics
}

By means of the formalism of paper \cite{Deco}, we will study a quantum
system with energy spectrum $0\leq \omega <\infty $ in the simplest case
where the CSCO is just $H$\footnote{%
More complete CSCO's are sudied in papers \cite{Deco} and \cite{Rolo}.}. The
observable $O$, belonging to some space ${\cal O}$, can be expressed in
terms of the Hamiltonian eigenbasis $\{|\omega \rangle \}$ as: 
\begin{equation}
O=\int O(\omega )\,|\omega )\,d\omega +\int \int O(\omega ,\omega ^{\prime
})\,|\omega ;\omega ^{\prime })\,d\omega \,d\omega ^{\prime }  \label{7.1}
\end{equation}
where $|\omega )=$ $|\omega \rangle \langle \omega |$, $|\omega ;\omega
^{\prime })=|\omega \rangle \langle \omega ^{\prime }|$, and $O(\omega
,\omega ^{\prime })$ are regular functions such that the Riemann-Lebesgue
theorem can be applied in eq.(\ref{7.1''}) \cite{LC}. Let us note that the
first term of the r.h.s. of the last equation is a diagonal operator, while
the second one is not diagonal. If $\rho $ is a quantum state belonging to a
convex ${\cal S\subset O}^{\prime }$, it can expanded as: 
\begin{equation}
\rho =\int \rho (\omega )\,(\omega |\,d\omega +\int \int \rho (\omega
,\omega ^{\prime })\,(\omega ;\omega ^{\prime }|\,d\omega \,d\omega ^{\prime
}  \label{7.1'}
\end{equation}
where $\left\{ (\omega |,(\omega ;\omega ^{\prime }|\right\} $ is the dual
basis of $\left\{ |\omega ),|\omega ;\omega ^{\prime })\right\} $. Again, $%
\rho (\omega ,\omega ^{\prime })$ are regular functions such that the
Riemann-Lebesgue theorem can be applied in eq.(\ref{7.1''}): the first term
of the r.h.s. of the last equation is a diagonal operator but not the second
one. Then, the time evolution of the mean value of $O$ in the state $\rho $, 
$\langle O\rangle _\rho $, reads: 
\begin{equation}
\langle O\rangle _\rho =(\rho |O)=\int \rho (\omega )\,O(\omega )\,d\omega
+\int \int \rho (\omega ,\omega ^{\prime })\,O(\omega ,\omega ^{\prime
})\,e^{-i(\omega -\omega ^{\prime })t\,}d\omega \,d\omega ^{\prime }
\label{7.1''}
\end{equation}
Using Riemann-Lebesgue theorem, we have: 
\begin{equation}
\lim_{t\rightarrow \pm \infty }\langle O\rangle _{\rho
(t)}=\lim_{t\rightarrow \pm \infty }(\rho |O)=\langle O\rangle _{\rho
_{*}}=(\rho _{*}|O)  \label{7.2}
\end{equation}
where $O$ is any observable of the space ${\cal O}$ and $\rho _{*}$ is a
diagonal operator: 
\begin{equation}
(\rho _{*}|=\int \rho _\omega \,(\omega |\,d\omega  \label{7.3}
\end{equation}
Then, we have the {\it weak limit}: 
\begin{equation}
W\lim_{t\rightarrow \pm \infty }(\rho |=(\rho _{*}|  \label{7.3'}
\end{equation}
and we obtain decoherence. This method is used in systems with more complex
energy spectra in papers \cite{Deco} and \cite{Rolo}.

i.- If the state $\rho $ at $t=0$ is a generic non-diagonal state, it {\it %
weakly} tends to the diagonal state $\rho _{*}$ towards the past and towards
the future, showing the time-reversal invariance of the underlying evolution
equations. This fact means that, if the normal equilibrium state of a local
system is essentially classical (namely, represented by a diagonal
operator), then it begins classical when $t\rightarrow -\infty $, is taken
out from this condition and becomes quantum when $t=0$ (that is, represented
by a non-diagonal density matrix), and finally tends again to a classical
state when $t\rightarrow +\infty $. Analogously to the case of Subsection
VII.D, this phenomenon can be decomposed in a pair of t-symmetric twins: the
process from $t=-\infty $ to $t=0$ and the process from $t=0$ to $t=\infty $%
. Again, the problem consists in supplying a non-conventional difference
between both processes, that is, in explaining why classicality emerges in
the process $t=0\rightarrow t=\infty $ but we never find a classical system
becoming quantum in the process $t=-\infty \rightarrow t=0$.

ii.-However, a system undergoing decoherence is just a particular case of an
unstable structure, which is created by pumping energy from the past and
which decays into equilibrium towards the future\footnote{%
The unstable nature of quantum systems is very well known in quantum
computation, where the system`s tendency to decohere is the main obstacle to
the implementation of information processing hardware that takes advantage
of superpositions.}. As we have seen, the Big Bang corresponds to high
energies and, therefore, it is most likely a quantum state that decays into
a classical universe radiating energy. This energy produces local quantum
states which, in turn, decay into classical states. In this way, the
universe, that begins in a quantum state, gradually becomes classical in
each one of its subsystem. This is the global description of the transition
from quantum mechanics to classical statistical mechanics in the universe.

Now we can go back to Subsection VIII.B in order to see how the notions of
that subsection are related with decoherence. The spontaneous evolution from
quantum to classical happens in $C_{+}(x)$ (as all the spontaneous
evolutions) and, therefore, it takes place only towards the future. If we
want to know the decoherence time (let us say, towards the future), we must
find the poles (introduced in the previous subsection) in the lower
half-plane of the complex energy plane and then consider the pole closer to
the real axis\footnote{%
Precisely, the poles of the Von Neuman-Liouville operator, which measure the
decaying of the non-diagonal components\cite{Rolo}.}: the inverse of its
imaginary part is the decoherence time. This means that decoherence towards
the future is related to the lower poles (since we are dealing with a
process towards classical equilibrium) and, therefore, to the outgoing
states. Symmetrically, the (anti)decoherence time towards the past is
related with the upper poles (since now we are dealing with a process that
takes the system out of classical equilibrium) and with incoming states. Of
course, since the poles are placed in symmetric positions, both decoherence
times are equal\footnote{%
They may be different if a time-reversal non-invariant interaction is
present, i.e., weak interaction.}. But lower poles are related with $%
C_{+}(x) $ and upper poles with $C_{-}(x)$. Therefore, the time-asymmetry of
the universe defines the decoherence direction since it tells us which are
the decaying poles (that define the decoherence time) and which are the
growing poles (that define the preparation time of the instabilities).

\subsubsection{Classical statistical mechanics$\rightarrow $classical
mechanics}

By means of the Wigner integral, in papers \cite{Deco}, \cite{IJTP} it is
shown how the classical statistical mechanical state, obtained by the
process explained above, is composed by a set of individual classical states
moving along classical trajectories in phase space\footnote{%
Therefore, it corresponds to a Frobenius-Perron evolution \cite{Mackey}.}.
In fact, the Wigner function $\rho _\omega ^W(q,p)$ of the quantum state $%
(\omega |$ of eq.(\ref{7.3}) reads: 
\begin{equation}
\rho _\omega ^W(q,p)=C\,\delta (\omega -H^W(q,p))\,\delta
(a_1-A_1^W(q,p))...\,\delta (a_N-A_N^W(q,p))  \label{7.4}
\end{equation}
where $C$ is a normalization constant, $H^W(q,p)$ is the classical
Hamiltonian, and $A_1^W(q,p)...A_N^W(q,p)$ are the Wigner functions of the
remaining commuting observables whose quantum numbers $a_1...a_N$ define the
state $(\omega |=(\omega ,a_1...a_N|$\footnote{%
In equations (\ref{7.1}) to (\ref{7.3}) we have only written the $\omega $
for conciseness.}. Therefore, $\rho _\omega ^W(q,p)$ corresponds to a
classical trajectory defined by the constants of motion: 
\begin{equation}
\omega =H^W(q,p)\qquad a_1=A_1^W(q,p)\qquad ...\ \qquad a_N=A_N^W(q,p)
\label{7.5}
\end{equation}
and the Wigner function $\rho _{*}^W(q,p)$ corresponding to $(\rho _{*}|$
reads: 
\begin{equation}
\rho _{*}^W(q,p)=\int \rho _\omega \,\rho _\omega ^W(q,p)\,d\omega
\label{7.6}
\end{equation}
This means that $\rho _{*}^W(q,p)$ is a classical statistical state
corresponding to the motions of the classical trajectories $\omega
=H^W(q,p), $ $a_1=A_1^W(q,p),...,a_N=A_N^W(q,p)$ weighted by the
probabilities $\rho _\omega $. To the extent that we can conceive a quantum
state as a quantum ensemble\footnote{%
A quantum state $\rho $ can always be conceived as a set of individual
quantum states since, being selfadjoint, it reads: 
\[
\rho =\sum_ip_i\,|i\rangle \langle i|\quad \qquad 0\leq p_i<1 
\]
Each $|i\rangle \langle i|$ is an individual state because, considered as a
matrix, it has probability 1 (certainty); in other words, it contains
maximum information since $Tr(|i\rangle \langle i|^2)=1$; therefore, if
conceived as an ensemble, it would represent a set of copies of the same
state $|i\rangle \langle i|$, thus, one $|i\rangle \langle i|$ is
sufficient. As a result, $\rho $ can be considered as a set of individual
states $|i\rangle \langle i|$ linearly combined in proportions $p_i$.}, the
quantum ensemble becomes a classical ensemble. Each classical individual
state (belonging to the classical ensemble) can be considered as a very
small region in phase space if the domain of its density function in phase
space is small, even though always satisfying the uncertainty principle $%
\Delta q\Delta p\eqslantgtr \hbar $ (we could imagine that it is a coherent
state). But if the action $S$ of the local system under consideration is
very large ($S>>\hbar )$, this small region becomes almost a geometrical
point that follows a classical trajectory for all physical effects related
with its size\footnote{%
This process can also be studied using Gell\'{}man-Hartle histories, as in
Appendix C of paper \cite{Deco}.}. As we are now in the classical domain
(since in practice $\hbar \rightarrow 0$), we can repeat the argument of
Subsection VII.D for the explanation of the process $probability\rightarrow
fact$, but now starting from a quantum unstable state and going towards its
classical limit.

We have not exhausted the list of the arrows of time, but we have shown that
the most important ones agree with GAT\footnote{%
The arrow of time corresponding to the decay of quantum states is treated in
paper \cite{Castagnino}, and the psychological arrow of time in paper \cite
{CG}.}.

\section{Conclusion}

The panorama is not completely closed yet: weak interactions should be
included in this scenario\footnote{$K_0$ and $\overline{K}_0$ are not twins;
therefore, they introduce a completely different arrow of time, the weak
interaction arrow \cite{Sachs}, \cite{Ruso}.}. Most likely they alone will
give a complete local explanation of time-asymmetry. However, this fact
would not diminish the relevance of the also complete global explanation
given by cosmology. From its very beginning, theoretical physics has tried
to combine its different chapters in an unified formalism, and it is well
known that unifications have always produced great advances in physics.
Therefore, our future challenge will be to unify the weak-interactions
explanation with the cosmological explanation, instead of abandoning the
latter in favor of the former as many local-minded physicists insist.

As it is well known, there is never a last word in physics. Nevertheless, we
can provisionally conclude that the global definition of the arrow of time
has no serious faults and, therefore, it can be used as a solid basis for
studying other problems related with the time-asymmetry of the universe and
its sub-systems.

\section{Acknowledgments}

We are very grateful to the criticisms of Pierre Coulet, who inspired us to
write Section II, and to Gloria Dubner and Elsa Giacani for the footnote 13.
This paper is partially supported by CONICET and University of Buenos Aires.

\appendix

\section{Criticism to Schulman\'{}s argument}

Schulman\cite{Schulman} exhibits a model in which two weakly coupled systems
would maintain opposite running thermodynamic arrows of time. From this
model he concludes that regions of opposite running arrows of time at
stellar distances from us are possible. This possibility would represent a
counter-example to our position: Schulman's model would show that a universe
consisting in two weakly coupled sub-universes A and B can have two regional
arrows of time pointing to opposite directions.

Even though Schulman's argument sounds convincing at first sight, it becomes
implausible when analyzed from a cosmological viewpoint. In Schulman's
proposal, the low entropy extremities of the sub-universes A and B are
opposed, and both sub-universes evolve towards equilibrium in opposed time
directions. Let us consider two cases:

1.- The sub-universe A is bigger than the sub-universe B\footnote{%
To fix the ideas, we can say ''externally bigger''. The cases ''more or less
bigger'' or ''almost symmetric'' can be included in the coarse-grained
version of the corkscrew theorem, since in these cases there is only a small
difference with a time-symmetric model. Then, these solutions have small
measure.} (this situation is not considered by Schulman). If
time-orientation is defined by entropy increase, in this case the
time-orientation of the whole universe A$\cup $B will agree with the
time-orientation of A, and B will go from equilibrium to non-equilibrium.
Nevertheless, the behavior of B is neither strange not unnatural: since
there is a flux of energy which, according to the time-orientation adopted,
must be considered as a flux from A to B, then we can consider that it is
such energy what takes the sub-universe B out of equilibrium. In other
words, the decreasing entropy of the {\it open} sub-universe B has the same
explanation as the decreasing entropy in the usual open systems that we find
in our everyday life.

2.- The sub-universe A is equal to B (the situation studied by Schulman,
where A and B are identical). In this case, the universe A$\cup $B is
perfectly time-symmetric. But, as the theorem of Section V has shown,
time-symmetry has vanishing measure: it requires an overwhelmingly
improbable fine-tuning of all the state variables of the universe.

But even in the time-symmetric case, it is not admissible to suppose that
the sub-universes A and B have opposite time-orientations. When considered
as a cosmological model, Schulman's simple model describes a time-orientable
universe. In a time-orientable manifold, continuous time-like transport has
conceptual priority over any method of defining time-orientation. In other
words, Schulman's universe should have a light-cone structure such that, if
we continuously transport a future pointing vector from the point $x\in $A
along some curve to the point $y\in $B, the transported vector will fall
into the future lobe $C_{+}(y)$. This means that A's future cannot be
different than B's future: there is an only future for the whole manifold,
defined by its light-cone structure.

However, who prefers to insist on the attempt to use entropy increase for
defining time-orientation could appeal to the following strategy: to define
the future direction of time as the direction of the entropy increase, for
instance, in the sub-universe A, and then to establish the time-orientation
in the sub-universe B by means of continuous time-like transport. But who
adopts this strategy is committed to explain why future is the direction of
entropy increase in one region of the universe but not in the other: why the
entropy definition works in one region of the universe but not in all of
them. These considerations lead us to our starting point: the problem of the
arrow of time should be addressed from a global perspective, taking into
account the geometrical properties of space-time.


\begin{references}
\bibitem{Cosmo}  M. Castagnino, F. Gaioli and E. Gunzig, {\it Found. Cosmic
Phys.}, {\bf 16, }221, 1996.

\bibitem{CG}  M. Castagnino and E. Gunzig, {\it Int. Jour. Theo. Phys}., 
{\bf 36}, 2545, 1997.

\bibitem{Castagnino}  M. Castagnino and R. Laura, {\it Phys. Rev. A}, {\bf 56%
}, 108, 1997.

\bibitem{Goslar}  M. Castagnino, ''The global nature of time asymmetry and
the Bohm-Reichenbach diagram''{\it , }in A. Bohm, H. Doebner and P.
Kielarnowski (eds), {\it Irreversibility and Causality (Proc. G. 21 Goslar
1996)}, Springer-Verlag, Berlin, 1998.

\bibitem{Peyresq}  M. Castagnino and E. Gunzig, {\it Int. Journ. Theo. Phys}%
. {\bf 38,} 47, 1999.

\bibitem{Ordo}  M. Castagnino, J. Gueron and A. Ordo\~nez, {\it J. Math. Phys%
}., {\bf 43}, 705, 2002.

\bibitem{cqg1}  M. Castagnino and C. Laciana, {\it Class. Quant. Grav.}, 
{\bf 19,} 2657, 2002.

\bibitem{cqg2}  M. Castagnino, G. Catren and R. Ferraro, ''Time asymmetries
in quantum cosmology and the searching for boundary conditions to the
Wheeler-De Witt equation'', {\it Class. Quant. Grav}., in press, 2002.

\bibitem{Boltz}  L. Boltzmann, {\it Ann. Phys}., {\bf 60, }392, 1897.

\bibitem{Feynman}  R. P. Feynman, R. B. Leighton and M. Sands, {\it The
Feynman Lectures on Physics, Vol. 1, }Addison-Wesley{\it , }New York, 1964.

\bibitem{Davies}  P. C. Davies, ''Stirring Up Trouble'', in J. J. Halliwell,
J. Perez-Mercader and W. H. Zurek (eds.), {\it Physical Origins of Time
Asymmetry}, Cambridge University Press, Cambridge, 1994.

\bibitem{Ear}  J. Earman, {\it Phil. Scie}., {\bf 41, }15, 1974.

\bibitem{Schulman}  L. S. Schulman, {\it Phys. Rev. Lett}., {\bf 83,} 5419,
1999.

\bibitem{Penrose}  R. Penrose, ''Singularities and Time Asymmetry'', in S.
Hawking and W. Israel (eds.), {\it General Relativity, an Einstein Centenary
Survey}, Cambridge University Press, Cambridge, 1979.

\bibitem{Sachs}  R. G. Sachs, {\it The Physics of Time-Reversal}, University
of Chicago Press, Chicago, 1987.

\bibitem{HP}  H. Price, {\it Time's Arrow and the Archimedes' Point, }Oxford
University Press, Oxford, 1996.

\bibitem{Grun}  A. Gr\"unbaum, {\it Philosophical Problems of Space and
Time, }2nd. ed., Dordrecht, Reidel, 1973.

\bibitem{Matt}  G. Mattews, {\it Phil. Scie}., {\bf 46, }82, 1979.

\bibitem{Reichenbach}  H. Reichenbach, {\it The Direction of Time, }%
University of California Press, Berkeley, 1956.

\bibitem{HE}  S. Hawking and J. Ellis, {\it The Large Scale Structure of
Space-Time, }Cambridge University Press, Cambridge, 1973.

\bibitem{Lara}  M. Castagnino, H. Giacomini and L. Lara, {\it Phys. Rev. D}, 
{\bf 61}, 107302, 2000; {\bf 63}, 044003, 2001.

\bibitem{SH}  S. Hawking, ''Quantum cosmology and time asymmetry'', in J. J.
Halliwell, J. Perez-Mercader and W. H. Zurek (eds.), {\it Physical Origins
of Time Asymmetry}, Cambridge University Press, Cambridge, 1994.

\bibitem{PT}  M. Castagnino, {\it Phys. Rev. D}, {\bf 29, }2216, 1989. M.
Castagnino and F. D. Mazzitelli, {\it Phys. Rev. D}, {\bf 42, }482, 1990. M.
Castagnino and F. Lombardo, {\it Phys. Rev. D}, {\bf 48, }1722, 1993.

\bibitem{Lichne}  A. Lichnerowicz, {\it Th\'eories Relativistes de la
Gravitation et de l'Electromagn\'etisme}, Masson, Paris, 1955.

\bibitem{MV}  M. Visser, {\it Lorentzian Wormholes}, Springer-Verlag,
Berlin, 1996.

\bibitem{BV}  C. Barcel\'o and M. Visser, ''Twilight for the energy
conditions?'',{\it \ lanl.arXiv}, gr-qc/0205066, 2002.

\bibitem{LL}  L. Landau and E. Lifchitz, {\it Th\'eorie des Champs},
Editions Mir, Moscow, 1970.

\bibitem{CastKL}  M. Castagnino, M. Gadella, F. Gaioli and R. Laura, {\it %
Int. Jour. Theo. Phys}., {\bf 38}, 2823, 1999.

\bibitem{Aqui}  R. Aquilano, M. Castagnino and E. Eiroa, {\it Phys. Rev. D}, 
{\bf 59, }\#087301, 1999.

\bibitem{AV}  A. Vilenkin, {\it Phys. Rev. D}, {\bf 37, }888 , 1988.

\bibitem{MAC}  M. Castagnino, {\it Phys. Rev. D}, {\bf 57}, 750, 1998.

\bibitem{Dubner}  J.V. Narlikar, {\it From Black Clouds to Black Holes, }%
World Scientific, Singapore, 1995. D. Clark, {\it Supernovae,} J. M. Dent \&
Sons, London, 1979.

\bibitem{MdU}  F. Adams and G. Laughlin, {\it Rev. Mod. Phys.}, {\bf 69,}
337, 1997.

\bibitem{BG}  L. Bergstr\"om and A. Goobar, {\it Cosmology and Particle
Astrophysics}, John{\it \ }Wiley \& Sons, New York, 1999.

\bibitem{LP}  P. D. Lax and R. S. Phillips,{\it \ Scattering Theory},
Academic Press, New York, 1979.

\bibitem{Zeh}  H. D. Zeh. {\it The Physical Basis of the Direction of Time},
Springer, Berlin, 1989.

\bibitem{Mackey}  M. C. Mackey, {\it Rev. Mod. Phys}., {\bf 61}, 981, 1989.

\bibitem{Ehrenfest}  P. Ehrenfest and T. Ehrenfest, {\it The Conceptual
Foundations of the Statistical Approach in Mechanics}, Cornell University
Press, Ithaca, 1959 (original 1912).

\bibitem{Ruso}  L. S. Schulman, {\it Time's Arrow and Quantum Measurements,}
Cambridge University Press, Cambridge, 1997. A. I. Ajiezer, {\it M\'etodos
de la F\'\i sica Estad\'\i stica}, Mir, Moscow, 1981.

\bibitem{LBC}  F. Lombardo, L. Bombelli and M. Castagnino, {\it Jour. Math.
Phys.}, {\bf 39, }6040, 1998.

\bibitem{Ludwig}  G. Ludwig, {\it An Axiomatic Basis of Quantum Mechanics, }%
Springler, Berlin, 1987.

\bibitem{BohmPR}  A. Bohm,{\it \ Quantum Mechanics, Foundations and
Applications}, Springer-Verlag, Berlin 1979. A. Bohm, M. Gadella, {\it Dirac
Kets, Gamow Vectors, and Gel'fand triplets}, Springer- Verlag, Berlin, 1989.

\bibitem{Sudarshan}  C. G. Sudarshan, C. B. Chiu and V. Gorini, {\it Phys.
Rev. D}, {\bf 18, }2914, 1978.

\bibitem{LC}  R. Laura and M. Castagnino, {\it Phys. Rev A}, {\bf 57}, 4140,
1998. R. Laura and M. Castagnino, {\it Phys. Rev. E}, {\bf 57}, 3948, 1998.

\bibitem{Deco}  M. Castagnino and R. Laura, {\it Phys. Rev. A}, {\bf 62,}
\#022107, 2000.

\bibitem{Rolo}  R. Laura, M. Castagnino and R. Id Betan, {\it Physica A}, 
{\bf 271, }357, 1999.

\bibitem{IJTP}  M. Castagnino and O. Lombardi, ''The Self-Induced Approach
to Decoherence in Cosmology'', {\it Int. Jour. Theo. Phys}., in press, 2002.
\end{references}
\end{document}